\documentclass[]{aastex631}
\usepackage{amsmath}
\usepackage{graphicx}
\usepackage{threeparttable}
\usepackage{txfonts}
\usepackage{soul}

\begin{document}
\title{A Unified Volumetric Rate–Energy Relation from Magnetar Radio Bursts to Fast Radio Bursts}

\correspondingauthor{Fa-Yin Wang}
\email{fayinwang@nju.edu.cn}

\author[0009-0006-4945-4734]{Hao-Tian Lan}
\affiliation{School of Astronomy and Space Science, Nanjing University Nanjing 210023, China}
\affiliation{Purple Mountain Observatory, Chinese Academy of Sciences, Nanjing 210023, China}

\author[0000-0003-0672-5646]{Shuang-Xi Yi}
\affiliation{School of Physics and Physical Engineering, Qufu Normal University, Qufu 273165, China}

\author[0000-0003-4157-7714]{Fa-Yin Wang}
\affiliation{School of Astronomy and Space Science, Nanjing University Nanjing 210023, China}
\affiliation{Purple Mountain Observatory, Chinese Academy of Sciences, Nanjing 210023, China}
\affiliation{Key Laboratory of Modern Astronomy and Astrophysics (Nanjing University) Ministry of Education, China}

\begin{abstract} 
Fast radio bursts (FRBs) are millisecond radio pulses with extremely high bright temperature. Their physical origin is still a mystery. 
The discovery of FRB 20020428 supports the idea that at least a portion of FRBs is generated by magnetars. However, FRB 20200428 and other radio bursts of SGR 1935+2154 are much less energetic than that of extragalactic FRBs. Thus, whether the progenitors of extragalactic FRBs are magnetars is still in controversy. 
Here, we investigate the volumetric rates of radio bursts from SGR 1935+2154, non-repeating FRBs and FRB 20180916B using the uniform samples detected by the Canadian Hydrogen Intensity Mapping Experiment (CHIME). 
We find that they share a similar relation between the volumetric rate $R$ and the burst energy $E$, i.e., $R\propto E^{-\gamma}$ with $\gamma=1.31\pm 0.13$ from $10^{29}$ erg to $10^{42}$ erg. 
Our results support the hypothesis that both repeating and non-repeating FRBs originate from magnetars.
\end{abstract}

\section{Introduction}\label{sec1}
Fast radio bursts (FRBs) are transient radio pulses lasting a few milliseconds and are believed to originate from distant galaxies \citep{Cordes2019,Zhang2023,Wu2024}. More than 800 FRB sources have been detected \citep{CHIME/FRBCollaboration2019,Chime/FrbCollaboration2023}. Since the discovery of FRBs, their physical origin has raised essential questions which are widely discussed. Various models are proposed to explain the physical origin of them \citep{Platts2019}. For example, the ``close-in" model suggests that FRBs originate near or within the neutron star magnetosphere \citep{Yang2018,Kumar2020,Lu2020}, while the ``far-away" model proposes that FRBs are generated far from the central engine \citep{Beloborodov2017,Metzger2019}.  The leading source model for FRBs is a compact object \citep{Platts2019,Katz2020,Sridhar2021}. However, the nature of FRB sources is not fully understood. 

In 2020, FRB 20200428 was localized in the direction of SGR 1935+2154 in the Milky Way \citep{CHIME/FRBCollaboration2020b,Bochenek2020}. This supports the hypothesis that magnetars can power FRBs. However, the energy of FRB 20200428 is only about $10^{35}$ erg \citep{Wu2020}, which is significantly lower than that of extragalactic FRBs mainly ranging from $10^{37}$ to $10^{43}$ erg. Follow-up observations by CHIME have detected additional five radio bursts with energies from $10^{29}$ erg to $10^{32}$ erg, approximately ten orders of magnitude less energetic than those from distant galaxies \citep{Giri2023}. Unlike FRB 20200428’s isolated magnetar origin, rotation measure observations indicate that the sources of some repeating FRBs may reside in a binary system \citep{Wang2022,Anna-Thomas2023,Li2025}. Thus, whether the physical origin of FRB 20200428 is the same as that of FRBs from distant galaxies remains uncertain.
Another important question is whether repeating and non-repeating FRBs have the same progenitors. \cite{Pleunis2021} found that the repeating FRBs have larger width and narrower bandwidth compared with non-repeating ones for the CHIME catalog 1. Further studies of burst morphology with higher time resolution in the CHIME/FRB baseband data show similar results \citep{Sand2024}. 

A natural way to solve the above problems is to study the volumetric rate of FRBs \citep{Ravi2019,Lu2019,Zhang2020,Lu2022}. Previous studies mainly focus on how the population of FRBs evolves with redshift. \cite{James2022} analyzed non-repeating FRBs observed by Australian Square Kilometre Array Pathfinder (ASKAP), and found that the evolution of the FRB population with the redshift scales with the star formation rate (SFR). \cite{Hashimoto2022} found that the volumetric rate decreases towards higher redshift which follows the evolution of the stellar mass density. Using the Lynden-Bell's c$^{-}$ method, \cite{Chen2024} concluded that the FRB rate is related to old populations, similar as short gamma-ray bursts. This result is supported by the following analysis \citep{Zhang2024,Champati2025,Zhou2025}. Other studies indicate that the FRB population has a time delay with respect to the star formation history \citep{Lin2024,2025arXiv250109810G}.

Here, we focus on how the population of FRBs evolves with burst energy to study the relation between the radio bursts generated by SGR 1935+2154 and extragalactic FRBs observed by CHIME. Studying the volumetric rate of FRBs with burst energy faces a significant challenge from the selection effect and observational strategy. For example, FRBs observed by different telescopes will face varying observational conditions and sensitives. Furthermore, the selection strategy of the observational target for one telescope also leads to selection bias. Fortunately, CHIME has a uniform observational mode, which can eliminate the selection effect of observational targets \citep{CHIME/FRBCollaboration2019}. 

This Letter is organized as follows. In Section~\ref{sec1}, we introduce the data sample used in the analysis. In Section~\ref{sec2}, we describe the method used to calculate the volumetric rate in different energy ranges. In Section~\ref{sec3}, we fit the volumetric rate for radio bursts of SGR 1935+2154, non-repeating FRBs and FRB 20180916B. Finally, the discussion and conclusions are presented in Section~\ref{sec4}. 

\section{Data sample}\label{sec1}
CHIME is sensitive to the frequency range of 400-800 MHz \citep{CHIME/FRBCollaboration2019}. The CHIME detector observes the sky as it transits each day. Its field of view is about 256 deg$^2$. This observational mode can remove the selection effect caused by the selection of the target, providing a significant advantage for studying the volumetric rate of FRBs. 

First, in order to study the relation between the extragalactic non-repeating FRBs and the radio bursts of SGR 1935+2154, we consider the non-repeating FRBs in CHIME catalog 1 and the radio bursts emitted by SGR 1935+2154 observed by CHIME. CHIME catalog 1 contains 536 bursts observed by CHIME telescope from 2018 July to 2019 July. Among them, 474 bursts are non-repeating FRBs and the remaining 62 bursts come from 18 repeating FRBs \citep{CHIME/FRBCollaboration2021}. For SGR 1935+2154, CHIME observed a total of 6 bursts from August 28, 2018 to December 1, 2022 \citep{CHIME/FRBCollaboration2020b,Giri2023}. Among the six detections, three bursts were observed during the transit of the source in the main lobe of CHIME with a cumulative duration of 111.4 hours and a $95\%$ confidence level fluence threshold of $10.3$ Jy ms. The other three bursts were detected in the side-lobes with an exposure time of 627 days above a fluence threshold of $10.2$ kJy ms, assuming a Poisson process at $95 \%$ confidence level \citep{Giri2023}. To further examine the relation between repeating and non-repeating FRBs, we also include the bursts of FRB 20180916B. This consists of 38 bursts observed by CHIME from August 28, 2018 to September 30, 2019, with a total exposure time of 64 hours and a fluence threshold of 5.2 Jy ms at $90 \%$ confidence level \citep{Chime/FrbCollaboration2020a}. 

\section{Methods}\label{sec2}
\subsection{Redshift and Energy}
\begin{figure}
	\centering
	\includegraphics[width=0.6\columnwidth]{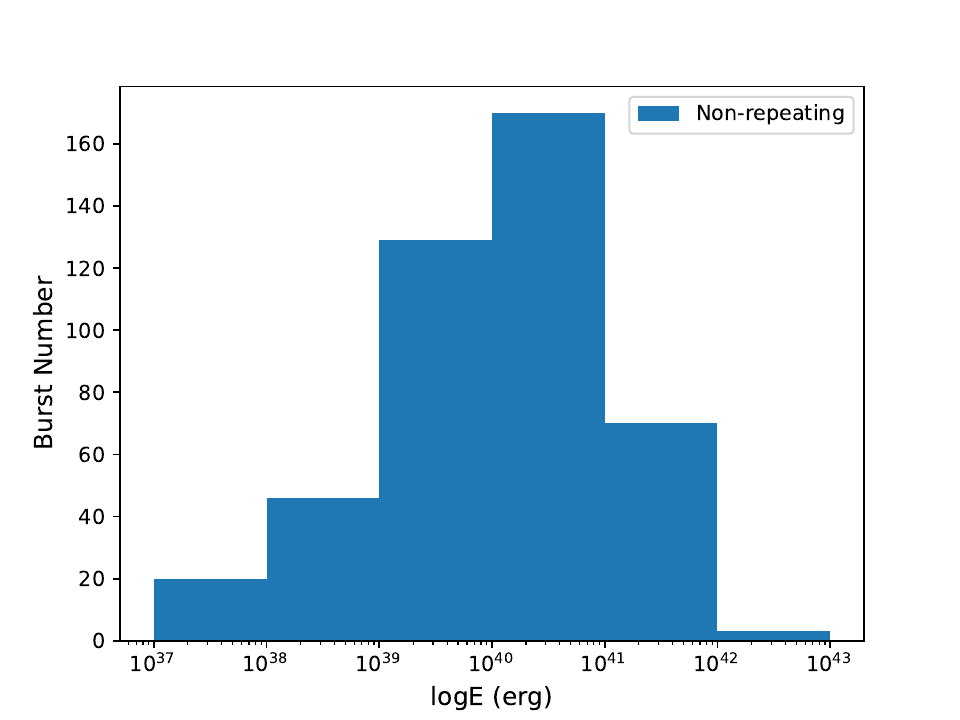}
    \includegraphics[width=0.6\columnwidth]{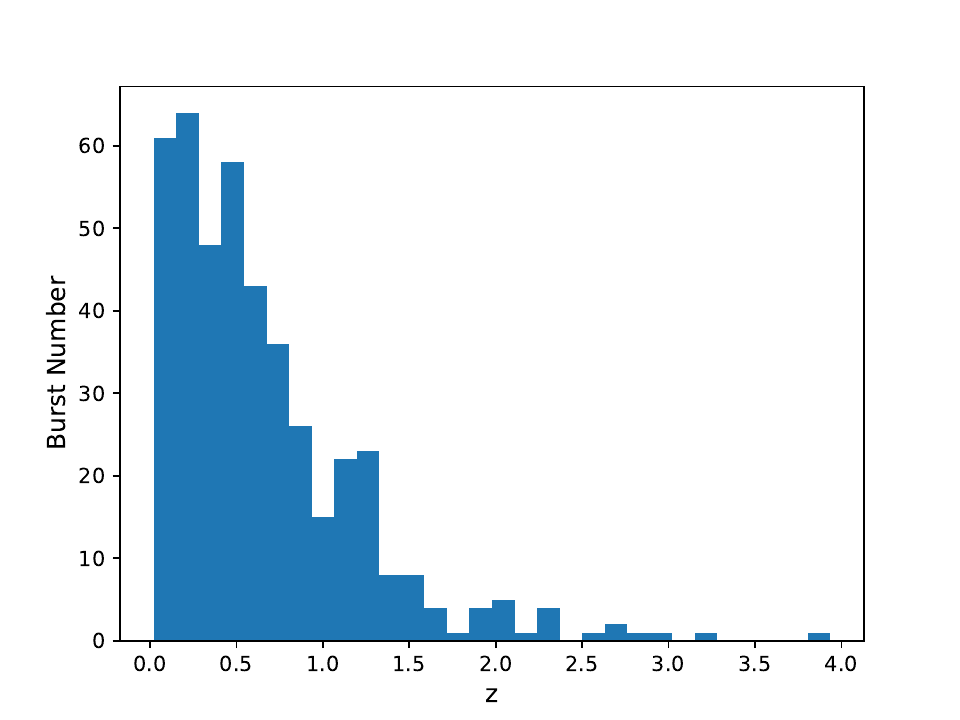}
	\caption{The energy and redshift distributions of the non-repeating FRBs in the CHIME Catalog 1. The upper panel shows the energy distribution of the non-repeating FRBs. The bottom panel gives the redshift distribution of the non-repeating FRBs.}\label{fig1}
\end{figure}

The redshift of FRB 20180916B is determined by its host galaxy localization \citep{Marcote2020}. The distance of FRB 20200428 is about 9 kpc. For the non-repeating FRBs in the CHIME Catalog 1, where precise localization is unavailable, their redshifts are estimated using the observed dispersion measure (DM), following the method described in \cite{Tang2023}. The effect of the redshift uncertainty on the final results is discussed in Section 4. The redshift distribution is shown in the top panel of Fig.~\ref{fig1}. The redshifts of these non-repeating FRBs range $0.02$ to $4.0$. The number of FRBs decreases with the increasing of redshift. 
The isotropic energy of each burst can be determined from
\begin{equation}\label{Eq_1}
    E=\frac{4\pi D_{L}^2(z)F_v\Delta v}{(1+z)^{2+\alpha}},
\end{equation}
where $D_{L}(z)$ is the luminosity distance at redshift $z$, $F_v$ is the fluence of the burst, $\Delta v=400$ MHz is the bandwidth of the CHIME telescope and $\alpha$ is the spectral index ($F \propto v^{\alpha}$). Here, $\alpha$ is set as $-1.4$, which is the best-fitting result for non-repeating FRBs from CHIME Catalog 1 \citep{CHIME/FRBCollaboration2020b,Shin2023}. As shown in the bottom panel of Fig.~\ref{fig1}, the energy of these non-repeating FRBs ranges from $10^{37}$ erg to $10^{43}$ erg.

\subsection{Volumetric rate}\label{sec2.2}
The volumetric rate of FRBs with different energies also evolves with redshift. Previous works used a simple scaling with star formation rate (SFR) assumption to model this evolutionary trend \citep{James2022,Shin2023}
\begin{equation}\label{Eq_SFR}
    \Phi(z)=\frac{\Phi_0}{1+z}(\frac{\mathrm{SFR}(z)}{\mathrm{SFR}(0)})^{n},
\end{equation}
where $\Phi(z)$ represents the volumetric rate of FRBs above certain energy and SFR$(z)$ is taken from \cite{2014ARA&A..52..415M}.
However, many studies \citep{Hashimoto2022,Chen2024,Zhang2024,Champati2025}, including \cite{Shin2023}, found that the volumetric rate of FRBs does not track the SFR as it evolves with redshift. The modeling and trend of the volumetric rate with respect to redshift are still debated. The other consideration is the assumption that the energy and the redshift are not independent with each other in the volumetric rate, which was actually found by \cite{Zhang2024}. This might further introduce systematic error in the study of volumetric rate. Considering the above two reasons, we choose the $V_{\mathrm{max}}$ method to eliminate the effect of redshift and focus solely on the relationship between the volumetric rate and energy \citep{Schmidt1968}.

\begin{figure}
	\centering
	\includegraphics[width=0.6\columnwidth]{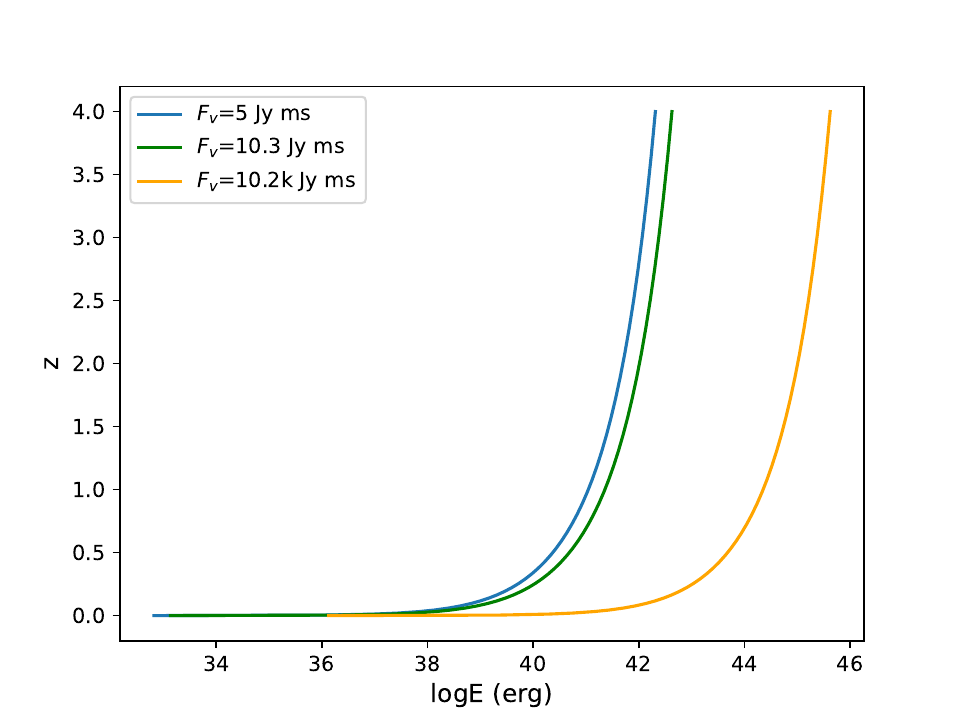}
	\caption{The detectable redshift as a function of burst energy from Eq.~(\ref{Eq_1}). Here, $\Delta v$ is set as $400$ MHz. The fluences $F_v$ of the orange, green and blue lines are $10.2$ kJy ms, $10.2$ Jy ms and $5$ Jy ms, which are the thresholds for the observations of SGR 1935+2154 in the side lobe, SGR 1935+2154 in the main lobe and  the FRBs in the CHIME Catalog 1, respectively.}\label{fig2}
\end{figure}

$V_{\mathrm{max}}$ method considers the varying energy completeness for the bursts with different energies. This is because the fluence of the detectable bursts has a lower limit. As shown in Fig.~\ref{fig2}, the relation between energy and its detectable redshift for a fixed fluence threshold is illustrated. It can be found that the detectable redshift range increases with higher energy for a fixed fluence threshold, while the redshift range decreases with a higher fluence threshold for a fixed energy. Thus, in this method, $V_{\mathrm{max}}$ represents the maximum comoving volume, within which a burst with a certain energy can be detected. It can be calculated by substituting the fluence threshold in Eq.~(\ref{Eq_1}). This method allows us to account for the completeness variations associated with different energies and fluence thresholds. It should be noted that, in principle, the fluence threshold is not a constant value, but rather a quantity that varies over time and across different beams. However, since this is a statistical analysis over several hundred days of observations, replacing the full convolution with a single fluence threshold corresponding to high completeness across all bursts, which is unlikely to introduce significant systematic errors. Therefore, in this work, we adopt the fixed fluence threshold provided by CHIME team \citep{CHIME/FRBCollaboration2020b,CHIME/FRBCollaboration2021,Giri2023}.

The volumetric rate for a given energy can be derived using $V_{\mathrm{max}}$ as
\begin{equation}\label{Eq_2}
    R=\frac{N(E)}{V_{\mathrm{max}}T_{\mathrm{obs}}\Omega_{\mathrm{sky}}},
\end{equation}
where $N(E)$ is the number of the bursts with energy $E$, and $\Omega_{\mathrm{sky}}$ is the fractional coverage of the CHIME field of the view on the sky. $T_{\mathrm{obs}}$ is the observational time of CHIME. 
In order to calculate the volumetric rate for different energy ranges, we divide the observational data into different energy bins. The 474 non-repeating FRBs are divided into 6 bins on a logarithmic scale along the energy axis from $10^{37}$ erg to $10^{43}$ erg. For the radio bursts of SGR 1935+2154, they are evenly divided into 4 bins on a logarithmic scale along the energy axis. Two bins correspond to energy ranges from $10^{29}$ erg to $10^{31}$ erg, observed by the main lobe. The other two bins correspond to energy ranges from $10^{32}$ erg to $10^{34}$ erg observed by the side lobe. The energy used to calculate the volumetric rate is set as the midpoint value of each energy bin. The uncertainties mainly arise from two sources: the Poisson fluctuations in the number of bursts within each energy bin, and the uncertainties along the horizontal axis. The error from the uncertainties along the horizontal axis for each bin can be calculated from
\begin{equation}\label{Eq_3}
    \sigma_{\mathrm{R,i}}=|\frac{dR(E)}{dE}|\sigma_{\mathrm{E,i}},
\end{equation}
where $\sigma_{\mathrm{E,i}}$ is the energy range of the $i$-th energy bin, and $dR(E)/dE$ is the derivative of the burst rate with respect to energy.

\section{Results}\label{sec3}
After calculating the volumetric rate with the observational data, the Bayesian method is employed to fit the relation between the volumetric rate and the burst energy \citep{Lu2019}. According to the Bayesian equation, the probability density function (PDF) of the posterior distribution for the parameters can be expressed as
\begin{equation}\label{Eq_4}
    f(p)\propto L(D\mid p)f_0(p),
\end{equation}
where $f(p)$ is the posterior probability density function (PDF), $f_0(p)$ is the prior distribution of the parameter which is set as the uniform distribution and $L(D\mid p)$ is the likelihood function of the observational data under certain parameters $p$. According to the least square method,
the logarithmic likelihood function can be written as
\begin{equation}\label{Eq_5}
    L=-\frac{1}{2}\sum_{i=1}^{N_{\mathrm{bins}}}[\frac{(R_{\mathrm{obs,i}}-R_{\mathrm{thre,i}})^2}{\sigma_{\mathrm{R,i}}}],
\end{equation}
where $N_{\mathrm{bins}}$ is the number of the energy bins, $R_{\mathrm{obs,i}}$ is the volumetric rate of the $i$-th energy bin derived from the observational data, and $R_{\mathrm{thre,i}}$ is the theoretical volumetric rate of the $i$-th energy bin.
 
\begin{figure}
	\centering
	\includegraphics[width=0.6\columnwidth]{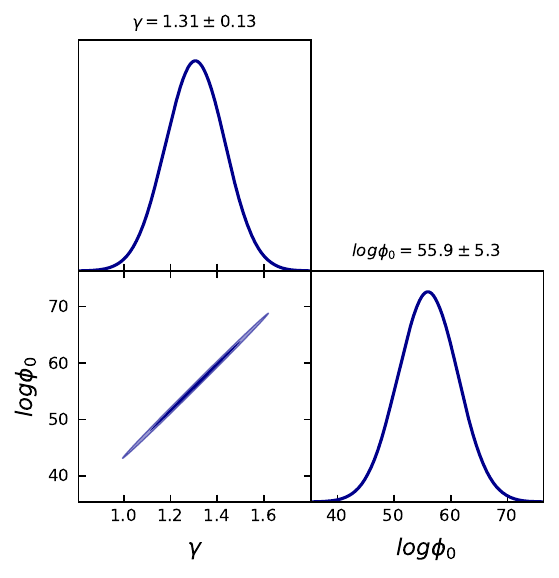}
	\caption{The contour plot of the posterior PDFs for parameters (log$\phi_0$, $\gamma$). The dark blue area represents the error range with $1\sigma$ confidence level.}\label{fig3}
\end{figure}

\begin{figure}
	\centering
	\includegraphics[width=0.8\columnwidth]{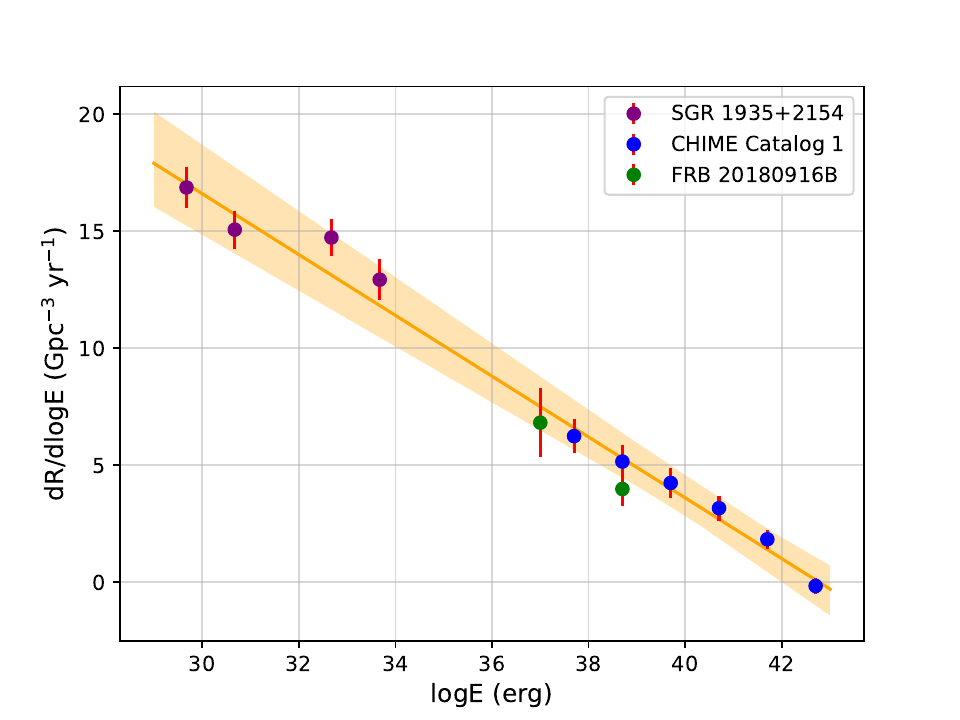}
	\caption{The volumetric rate as a function of burst energy. The purple, blue and green points are the volumetric rate of the radio bursts of SGR 1935+2154, non-repeating FRBs in the CHIME Catalog 1 and FRB 20180916B, respectively. The orange line is the best fitting result with \textbf{$\log\phi_0=55.9\pm 5.3$ and $\gamma=1.31\pm 0.13$}. The orange area is the error range of $1\sigma$ confidence level.}\label{fig4}
\end{figure}

The typical fitting function is considered as the Schechter function \citep{Lu2019,Lu2020,Hashimoto2022}
\begin{equation}\label{Eq_6}
    \frac{dR(E)}{dE}=\frac{\phi_0}{E_{\mathrm{max}}}(\frac{E}{E_{\mathrm{max}}})^{-\gamma}\mathrm{exp}(-\frac{E}{E_{\mathrm{max}}}),
\end{equation} 
where the $\phi_0$ is a constant, $\gamma$ is the energy power-law index and $E_{\mathrm{max}}$ is the maximal energy. The Schechter function initially decreases as a power-law function and it will decrease exponentially at $E=E_{\mathrm{max}}$. Previous results show that the cutoff occurs at approximately $10^{42}$ erg \citep{Lu2019,Lin2024}, which significantly exceeds the energy of the FRBs in CHIME Catalog 1. Our fitting result using the Schechter function yields a similar result. Thus, we consider the single power-law function
\begin{equation}\label{Eq_7}
    \frac{dR}{dE}=\phi_0E^{-\gamma},
\end{equation}
where $\phi_0$ is the constant and $\gamma$ is the power-law index. In this study, we only use the non-repeating FRBs to fit the power-law model. Similarly to the method above, we consider a uniform prior with $\gamma \in (0,5)$ and $\log\phi_0 \in (-5,300)$. The posterior PDF of the parameters are shown in Fig.~\ref{fig3}, where $\gamma=1.31\pm 0.13$. Using the fitting result, we plot the relation between the volumetric rate and the energy shown in Fig.~\ref{fig4}. Furthermore, we also use the non-repeating FRBs and the radio bursts from SGR 1935+2154 to fit the power-law model and get a similar result of $\gamma=1.322 \pm 0.045$ with a better constraint. The value is consistent with the finding of \cite{Shin2023}, who found a differential power-law energy distribution index of $1.3^{+0.7}_{-0.4}$ using CHIME Catalog 1. \cite{Lu2019} found a power-law index of $\gamma=1.6\pm 0.3$ from 20 non-repeating FRBs published by the ASKAP Fast Transient survey which is also aligned with our power-law index. In general, the value is generally consistent with those of previous works, with minor differences that could be due to the use of different datasets \citep{Lu2019,James2022,Hashimoto2022,Shin2023,2024Univ...10..207Z,2025arXiv250413705M,2025PASA...42....3A}. 

In addition to the selection effect mentioned above, variations in DM contributions from the interstellar medium of host galaxies and FRB sources \citep{Wang2022} can also introduce errors in the redshift estimation. This type of uncertainty is difficult to quantify and eliminate. Typically, this effect may be mitigated by the statistical effect of a large number of FRBs. However, we performed a refit using the redshift-DM relation from \cite{2024arXiv241003994G} to further verify our results. In calculating the DM-redshift relation, \cite{2024arXiv241003994G} used the probability distribution of DMs from FRB host galaxy \citep{2020ApJ...900..170Z} and intergalactic medium \citep{2021ApJ...906...49Z} derived from the IllustrisTNG simulation. The IllustrisTNG project is the successor of the Illustris project \citep{2018MNRAS.473.4077P}. It is a large-volume, cosmological, and magnetohydrodynamical simulation which can be used to derive the electron distribution. With use of this simulation, \cite{2020ApJ...900..170Z} and \cite{2021ApJ...906...49Z}  calculated the probability distributions of DMs contributed by host galaxy and intergalactic medium. Using the redshifts derived by \cite{2024arXiv241003994G}, we get a similar result of $\gamma=1.30 \pm 0.11$. This demonstrates that our result is weakly dependent on the method to derive the pseudo redshifts of FRBs.

The relation between the volumetric rate and energy can reflect their physical origin. This can be due to two reasons. Firstly, different FRB progenitors have different volumetric rates. Secondly, they may show different energy functions. 
In Fig.~\ref{fig4}, we also show the volumetric rate of repeating FRB 20180916B and the radio bursts from SGR 1935+2154 as a function of burst energy. We can see that the volumetric rate of FRB 20180916B falls on the power-law function of non-repeating FRBs. Moreover, the relation between volumetric rate and energy of the low-energy radio bursts generated by SGR 1935+2154 follows the extrapolation of the high-energy non-repeating FRBs.
This power-law relation holds across an energy range spanning thirteen orders of magnitude (10$^{29}$-10$^{42}$ erg), supporting that the physical origin of extragalactic non-repeating FRBs is the same as that of the radio bursts from SGR 1935+2154. Additionally, we also find that the volumetric rate of FRB 20180916B falls on the power-law function, which may indicate that the repeating and non-repeating FRBs have the same progenitor, i.e. magnetars. Consequently, non-repeating FRBs may be repeating FRBs with a lower burst rate. 

The volumetric rate of non-repeating FRBs also shows a valuable clue of their sources. The volumetric rate of the non-repeating FRBs in the local universe exceeds the rate of catastrophic events, and the birth rates of compact-object sources \citep{Ravi2019,Hashimoto2020}. This suggests that the non-repeating FRBs may actually repeat many times. However, their repetitions may have much less energy, which makes the repetitions without enough energy to be observed at extragalactic distance \citep{Katz2022}. 

Another problem is whether the discriminant properties between the repeating and non-repeating FRBs can also be explained in the framework of magnetars. The analysis of FRB properties confirms that the repeating FRBs and non-repeating FRBs are different in some aspects, such as repetition rate, burst width, bandwidth, and spectral index \citep{CHIME/FRBCollaboration2021,Pleunis2021}. Most of these discriminant properties can be efficiently explained by the selection effect due to the beamed emission \citep{Connor2020}. However, this explanation face problems in explaining the spectral index and potential peak frequency in the framework of the coherent curvature radiation or the synchrotron maser emission \citep{Zhong2022}. In order to solve this problem, a unified model was proposed considering the quasi-tangential propagation effect \citep{Liu2024}. In this model, the repeating and non-repeating FRBs are generated from magnetars with different emitting region, which the emitting region of the non-repeating FRBs has a smaller impact angle with respect to the magnetic axis than the repeating FRBs. This unified model can well explain the differences between the repeating and non-repeating FRBs, indicating they both generate from the magnetosphere. A promising trigger mechanism for FRBs is starquake on magnetars \citep{Wang2018,Suvorov2019}. For FRB 20200428, \cite{Yang2021} and \cite{Lu2020} suggested that the FRB and X-ray burst can be well explained if they are triggered by crust fracturing of magnetars. Further observations detected glitch events associated with radio bursts of SGR 1935+2154, supporting that radio bursts are triggered by starquakes \citep{Younes2023,Hu2024,Ge2024}. Direct evidence is that the energy functions of three active repeating FRBs show a universal break around $10^{38}$ erg, similar to that of earthquakes \citep{Wu2025}. Additionally, the non-detection of periodicity related to rotation can be explained by the rotation-modulated starquakes on magnetars \citep{Luo2025}. 

\section{Discussion and Conclusions}\label{sec4}


There are some studies on the burst rate of repeating FRBs, such as FRB 20201124A, FRB 20121102, and FRB 20190520B. \cite{Zhang2021} studied the burst rate of FRB 20121102 using 1,652 bursts observed by the Five-hundred-meter Aperture Spherical Telescope (FAST) \citep{Li2021} and found that, in the higher energy range (larger than $10^{38}$ erg), the differential energy distribution can be fitted with a single power-law function with an index of $\gamma=1.86\pm0.02$. \cite{Kirsten2024} utilized 46 bursts of FRB 20201124A detected by four 25–32 m class radio telescopes and found that the cumulative burst rate can be well fitted by a single power-law function with $\gamma=1.09\pm 0.03\pm 0.09$. A similar result $\gamma=1.54\pm 0.34$ is also derived from 35 bursts observed by the Allen Telescope Array \citep{Sheikh2024}. Recently, \cite{Wu2025} reported that the cumulative distribution of energy for three active repeating FRBs can be well fitted by a broken power-law model with a universal break around $10^{38}$ erg. However, it should be noted that these works discussed the energy (burst rate) distribution of a specific repeating FRB source, which is different from the volumetric rate discussed in this paper. These bursts are detected by purposeful observations. Therefore, the statistical results of the burst rate and the volumetric rate are not directly comparable. 

In conclusion, we investigate the volumetric rates of non-repeating FRBs in CHIME Catalog 1, FRB 20180916B, and radio bursts from SGR 1935+2154. Previous work mainly concentrated on the non-repeating FRBs in the distant galaxies with an energy range around 5 orders of magnitude \citep{James2022,Shin2023,2024Univ...10..207Z,2025arXiv250413705M,2025PASA...42....3A}. Here, we primarily focus on the relation among the non-repeating FRBs, FRB 20180916B and radio bursts from SGR 1935+2154 with an energy range over 13 orders of magnitude. We use the data observed by CHIME to eliminate the selection effect from different telescopes or observational modes. In addition, compared to previous studies that examined the relation between FRB 20200428 and non-repeating FRBs \citep{Lu2020,Bochenek2020}, we use six radio bursts from SGR 1935+2154 observed by CHIME. Their energies span about 4 orders of magnitude, which exhibits a slope consistent with that of non-repeating FRBs. Further, a larger sample of non-repeating bursts is used in our study. Both of the two reasons could significantly improve the confidence of this relation. Due to the observation mode of the CHIME telescope, we are also able to calculate the volumetric rate of FRB 20180916B. This allows us to further compare the relationship between FRB 20180916B, and non-repeating FRBs, as well as SGR 1935+2154. Finally, we find that these bursts conform to a single power-law function between volumetric rate and energy with a power-law index $\gamma=1.31 \pm 0.13$ across the energy range from $10^{29}$ erg to $10^{42}$ erg. This strong correlation over an energy range of 13 orders of magnitudes supports that the progenitors of extragalactic non-repeating FRBs and repeating FRBs are similar to the radio bursts from SGR 1935+2154, i.e. magnetars.

\section*{Acknowledgements}
We thank the two anonymous referees for constructive comments. We thank J. H. Chen for helpful discussion. This work was supported by the National Natural Science Foundation of China (grant Nos. 12494575 and 12273009) and the National SKA Program of China (grant No. 2022SKA0130100). We acknowledge the use of the CHIME/FRB Public Database, provided at https://www.chime-frb.ca/ by the CHIME/FRB Collaboration.

    \bibliographystyle{aasjournal}
    \bibliography{refs}

\end{document}